\documentstyle[aps,preprint,pre]{revtex}
\def\c#1{\setbox0=\hbox{#1}\ifdim\ht0=1ex\accent24 #1%
  \else{\ooalign{\hidewidth\char24\hidewidth\crcr\unhbox0}}\fi}
\begin{document}
\tightenlines
\author{
Jakub Zakrzewski$^{1,2}$, Dominique Delande$^1$,
  and Andreas Buchleitner$^3$}
\address{$^1$ Laboratoire Kastler-Brossel, Tour 12, Etage 1,
Universite Pierre et Marie Curie,\\
4 Place Jussieu, F-75005 Paris,\\
 $^2$ Instytut Fizyki imienia Mariana Smoluchowskiego,
  Uniwersytet Jagiello\'nski,\\
 ulica Reymonta 4, PL-30-059 Krak\'ow \footnote{permanent address},\\
$^3$ Max-Planck-Institut f\"ur Quantenoptik, Hans-Kopfermann-Str. 1,
D-85748 Garching.}
\title{Ionization via Chaos Assisted Tunneling}
\date{\today}
\maketitle
\begin{abstract}
A simple example of quantum transport in a classically chaotic
system is studied. It consists in a single state lying on a regular
island (a stable primary resonance island) which may tunnel into
a chaotic sea and further escape to infinity via 
chaotic diffusion.
The specific system is realistic : it is the hydrogen atom exposed
to either linearly or circularly polarized microwaves.
We show that the combination of tunneling followed by chaotic
diffusion leads to peculiar statistical fluctuation properties
of the energy and the ionization rate, especially to enhanced
fluctuations compared to the purely chaotic case.
An appropriate random matrix model, whose predictions are
analytically derived, describes accurately these
statistical properties.
\end{abstract}
\pacs{PACS: 05.45.+b, 32.80.Rm, 42.50.Hz}
\narrowtext
\section{Introduction}

Quantum systems with classically chaotic counterparts possess
unique characteristic features as summarized, e.g., in
\cite{rev,haake}. Following the semiclassical approach, one
often relates quantum properties of a system to its classical
motion, using for example a direct comparison of
 phase space portraits of the classical dynamics and
 wave function quasiprobability representations in the phase
space (via Husimi or Wigner functions) \cite{rev,rvj,holt}.
Even in the case of globally
chaotic dynamics, individual unstable classical trajectories
 can  be retraced
by stationary quantum eigenfunctions which are ``scarred" by the
classical
solution \cite{Hell84}. When the classical phase space is mixed -
partially chaotic and partially regular - a similar separation
into regular and irregular wavefunctions is possible in the
quantum world \cite{Perc73}.
 Stable regions of phase space (tori) lend
themselves to semiclassical EBK
 (Einstein-Brillouin-Keller)
quantization, yielding both the approximate eigenenergies and
the corresponding wavefunctions \cite{MF65}. Similarly,
 there are ``irregular" wavefunctions living in the region of
chaotic classical motion. Some of them can be associated
 with residual structures
of classically regular motion such as cantori while the other
are practically structureless.
In low-dimensional systems, KAM
(Kolmogorov-Arnold-Moser)
tori provide impenetrable
borders;
 the only way regular and
irregular wave functions may communicate with each other is by
quantum mechanical
tunneling processes. In higher-dimensional systems, classical Arnold
diffusion provides another mechanism of transport, a process
which is, however, typically quite slow \cite{LL81}. On the other
hand, quantum mechanical tunneling through impenetrable borders
of classical mechanics may be quite effective. Once the particle
tunnels from, say, a stable island into the surrounding chaotic
phase space, it can visit distant regions following the classically
chaotic transport mechanism. In particular, it can tunnel into some other
stable island thus providing the coupling between two
wavefunctions localized on distinct and separated islands or
it can wander very far away (possibly leading, e.g., to ionization
as in atoms driven by external radiation).

Interestingly, this 
``chaos assisted'' tunneling mechanism posseses
unique features typically absent in the standard ``barrier''
tunneling of quantum mechanics, such as a great sensitivity to
the variation of external parameters manifesting itself in
fluctuations of observable quantities. Previous work
considered mainly model one-dimensional time dependent
systems \cite{LB90,GDJH91,PGL92} or model two-dimensional
autonomous systems \cite{BTU93,BBEM93,TU94,LU96}. A similar
problem in the scattering case has also been discussed on a
kicked model system \cite{GS94}. We shall consider here a
realistic, experimentally accessible (although simplified, see
below) system - namely the hydrogen atom illuminated by
microwave radiation. Instead
of considering tunneling between two regions (tori) mediated by the
chaotic transport between them, we shall rather consider the
single tunneling process out of the stable island. Then the
chaotic diffusion process will lead to ionization. While in the
former case, the probability may flow periodically between two
regions linked by the tunneling coupling, in our problem the
process is irreversible and constitutes the mechanism of the
decay.

The paper is organized as follows. Section II contains the description
of the systems studied - the hydrogen atom in the field of a microwave
radiation of either circular or linear polarization - and a general
presentation of the ionization via chaos assisted tunneling.
Section III presents a simple
model for the description of the fluctuations present in the
decay, catalized by chaos assisted tunneling.
We present there the distribution of resonance widths
and consider also the distribution of energy shifts of the
single, initially localized state due to the coupling to other
``delocalized'' states, and, via these states, to the continuum.
This theory is confronted with
the numerical data obtained for the hydrogen atom in the field
of microwave radiation of either circular or linear polarization
in Section IV. Finally we present conclusions in Section V
while the Appendix contains the details of the derivation of
formulae presented in Section III.

\section{Nondispersive electronic wave packets and their ionization}

In order to obtain the simplest study of quantum transport
through a chaotic sea, one should use an initial state as localized
as possible in phase space as, for example, a minimum wave packet
localized on a classical stable equilibrium point.
Unfortunately, in atomic systems, no stable equilibrium point of the electron
outside the nucleus exists.

A simple alternative is to use a nonlinear resonance between the
internal motion of the electron and an external driving force.
Recently, interesting new objects have been proposed in the studies of
hydrogen atoms illuminated by microwave radiation of either linear \cite{ab1}
or circular \cite{bb} polarization: the so called non-dispersive
wave packets. The corresponding classical dynamics picture
corresponds to the stable  resonance island
embedded in a chaotic sea. For the motion contained within
the principal 1:1 resonance between the Kepler frequency of the
unperturbed Rydberg electron and the frequency of the driving field,
the frequency of the electronic motion is locked on the external
microwave frequency. Semiclassically, a wave packet localized on such a
regular island will be confined to it modulo 
the exponentially
decaying
tails of the wavefunction which may extend into the chaotic region.
In a quantum treatment, one finds
wave packets which are really single eigenstates of the atom dressed
by the microwave field, i.e. single eigenstates of the
corresponding Floquet \cite{sfloq} Hamiltonian \cite{ab1,dzb95}.
They are localized in all spatial dimensions and propagate along the classical
trajectory in the same way a classical particle would do.
For a generic case (e.g.
linear polarization microwaves, or more generally, any time-periodically
perturbed system \cite{holt}), it undergoes periodic deformations
which faithfully follow the change of shape of the resonance island
over one period, repeating the same shape every period.
Only in the case of circular microwave polarization,
the shape of the wave packet eigenstate does not change. This is
due to the fact that the time-dependence may be removed from the
Hamiltonian of the problem by a transformation to the frame
rotating with the field \cite{proh,groz,zgd96}.

As mentioned above, a finite
$\hbar$ value leads to quantum mechanical tunneling from
the island to the chaotic sea surrounding it. Then the electron
gains energy 
from the driving field 
and eventually becomes ionized by a process
classically known as chaotic diffusive ionization. Since many different
paths link the initial wave packet with the continuum, its
ionization time (or its reciprocal - the ionization rate or resonance width)
fluctuates strongly with the parameters of the problem, the
microwave frequency, $\omega$, or its amplitude, $F$ 
\cite{zdb95}.
Therefore, these wave packets are ideally suited for a quantitative
study of the ionization promoted by chaos assisted tunneling.

\subsection{Circularly polarized microwaves}

Let us consider first the conceptually simpler case of hydrogen atoms
illuminated by a circularly polarized microwave (CPM) \cite{bb,dzb95,zdb95}.
The problem is fully three dimensional; however, as
it has been shown elsewhere \cite{dzb95,zdb95}, one can consider
the quantum dynamics
in the space  restricted to the polarization plane of the microwave field.
While
this excludes possible excitations in the direction
perpendicular to the polarization plane, the dynamics of
the wave packets and their properties are qualitatively
not affected by the reduced dimensionality \cite{dzb95,zdb95}. In the
following, we shall present results obtained within
such a reduced two-dimensional (2D) model.

The time-dependence of the
Hamiltonian describing the CPM ionization of H atoms is explicitely
removed by transforming to the coordinate frame which
corotates with the microwave field \cite{proh,groz},
where it reads (in atomic units):
\begin{equation}
H=\frac{{\bf p}^2}{2}-\frac{1}{r}+Fx-\omega\ell_z,
\label{hcirc}
\end{equation}
with $\ell_z$ the angular momentum operator.

At the center of the 
principal resonance island between the Kepler
and the microwave frequency, a periodic orbit exists whose period
exactly matches
the period of the microwave. In the laboratory frame, this
is a circular orbit with radius $x$ such that:
\begin{equation}
\frac{1}{\omega^2x^2}+\frac{F}{\omega^2}=x.
\label{pos}
\end{equation}
We introduce the effective
principal quantum number $n_0$ (not necessarily an integer) corresponding
to this main resonance:
\begin{equation}
n_0 = \omega^{-1/3}.
\label{n0}
\end{equation}
Due to the classical scaling of the Coulomb problem \cite{del},
among the two parameters
$\omega$ and $F$
only one is necessary to 
tune the dynamics classically.
Thus, we define quantities (position and microwave electric field) scaled
with respect to $n_0$:
\begin{equation}
\left\{
\begin{array}{l}
x_0 = x n_0^{-2} = x \omega^{2/3}\\
F_0 = F n_0^{4} = F \omega^{-4/3}
\end{array}
\right.
\end{equation}
$F_0$ represents the ratio of the external microwave
field to the Coulomb field of the nucleus on the
unpertubed resonant circular orbit. The classical dynamics
depends only on this parameter.
The scaled radius $x_0$ of the resonant circular orbit is the solution
of the scaled equation:
\begin{equation}
\frac{1}{x_0^2}+F_0=x_0
\label{pos0}
\end{equation}

In the corotating frame, the resonant orbit corresponds to an equilibrium point.
This point is stable if the
dimensionless stability parameter
\begin{equation}
q=\frac{1}{\omega^2x^3}=\frac{1}{x_0^3}
\label{stab}
\end{equation}
is chosen in the interval $8/9<q<1$ \cite{bb}. Then, the existence of a
wave packet localized in the vicinity of the fixed point is assured in
the semiclassical limit. It appears in the rotating frame as a
stationary eigenstate of the Hamiltonian, Eq.~(\ref{hcirc}), localized
around the equilibrium point and in the laboratory frame as a localized
wave packet following a circular trajectory without spreading.

The energies of the wave packet eigenstates in the
rotating frame are given (in 
harmonic approximation around the stable fixed point) by \cite{dzb95}
\begin{equation}
E_0=\frac{1-4q}{2q^{2/3}}\omega^{2/3}+(N_++\frac{1}{2})\omega_+-(N_-
+\frac{1}{2})\omega_-,
\label{semicl}
\end{equation}
with
\begin{equation}
\omega_{\pm}=\sqrt{\frac{2-q\pm\sqrt{q(9q-8)}}{2}}\omega
\label{modes}
\end{equation}
the normal mode frequencies of the locally harmonic Hamiltonian which confines
the wave packet, and $N_{\pm}$ the number of quanta in these modes.
In the following, we shall consider the ionization of the ``ground
state'' wave packet corresponding to $N_+=N_-=0$.
Such a wave packet can be expanded over the usual atomic eigenstates.
It is a coherent superposition of mainly circular
states \cite{dzb95}. The frequency  $\omega $ is
close to the resonance between atomic circular states with
principal quantum numbers
$n\rightarrow n\pm 1$ with $n\simeq n_0$.
Thus these states are strongly coupled by the
microwave field. It can be shown that, for such a frequency, the overlap
of the wave packet state with circular states is, for a sufficiently high
$n_0,$ Gaussian distributed with a maximum at $n_0$ and the width
of the order of $\sqrt{n_0}$.

To find the wave packets numerically, we diagonalize the time-independent
Hamiltonian in the rotating frame,
Eq.(\ref{hcirc}), in a Sturmian basis \cite{zgd96}.
The so-called complex rotation method
\cite{rein}
allows us to take exactly into account the coupling to the continuum.
We refer the reader to Ref.~\cite{zgd96}
for the description of the technical details. Let us mention here only
that a diagonalization yields complex energies, $E_i-i\Gamma_i/2$, where
the real parts
$E_i$ are the positions of the resonances while the $\Gamma_i$ correspond
to their widths, i.e. their ionization rates.
In this approach, spontaneous emission from the wave packet eigenstates
to lower lying states is neglected. This is a reasonable approximation
as they are composed of mainly circular states which have very long
spontaneous lifetimes, typically of the order of several millions periods.
In all calculations discussed hereafter, the decay of wave packet
eigenstates is dominated by field induced ionization (via chaos assisted
tunneling) and {\em not} by spontaneous emission \cite{ab97,ibb97}.

The present results are obtained from the
diagonalization of matrices of size up to $200000$. We
use the Lanczos algorithm \cite{gd91} to extract few eigenvalues and
the corresponding eigenvectors in the vicinity of the energy
predicted by the semiclassical expression, Eq.~(\ref{semicl}). The
wave packet eigenstate is then identified by its large
overlap with the circular state with principal quantum number 
close to $n_0=\omega^{-1/3}$, and by its large dipole moment. Typically, due
to the accuracy of the semiclassical prediction, it is enough
to extract 
few eigenvalues of the matrix. For 
our present purpose, it is the
{\it deviation} of the exact resonance position from the semiclassical
prediction, and the {\it ionization rate}
which are of great interest for us.

For our statistical analysis, it is
reasonable to collect the data,
for a {\it  fixed} classical dynamics, i.e. at {\it fixed} $F_0$
value varying simultaneously
$\omega$ and $F$ around some mean values.
On the other hand, quantum mechanics does {\em not} preserve the classical
scaling, the finite $\hbar$ value introduces another scale into the
problem - say the energy of the state or
 the number of photons necessary for the ionization. We shall see
that this has important consequences for the characteristics of the
ionization process -- the statistical properties of the widths
depend on $n_0$.

We shall present the numerical data for both circular and
linear polarization 
of the driving field simultaneously. Thus before presenting the data,
we review the wave packet properties in linearly polarized
microwaves (LPM)
\cite{ab1}.

\subsection{Linearly polarized microwaves}

If the atom is irradiated by an electromagnetic field of linear
polarization defining the $z$-axis, the angular momentum projection
$m$ on the field axis is a good quantum number, due to the rotational
symmetry of the Hamiltonian:
\begin{equation}
H=\frac{{\bf p}^2}{2}-\frac{1}{r}+Fz\cos(\omega t).
\label{hlp}
\end{equation}
One is therefore left with two spatial degrees of freedom and the
explicit, time periodic dependence of the Hamiltonian which cannot be
eliminated. However, the temporal periocicity of the problem (for
constant driving field amplitude $F$) allows
for the application of the
Floquet theorem and the identification of the eigenfunctions
$|\psi_i \rangle$ of the
atom {\em in} the field as solutions of the stationary Floquet
equation \cite{sfloq}
\begin{equation}
(H-i\partial _t)|\psi_i \rangle=\epsilon|\psi_i \rangle,
\label{floqeq}
\end{equation}
where spatial coordinates and time are treated now on an equal
footing. The Floquet theorem guarantees that the eigenfunctions
$|\psi_i \rangle$ are periodic with the period $T=2\pi/\omega$ of the
driving field, and form a complete basis of the problem.
The Floquet states are nothing but the ``dressed
states'' of the atom in the field \cite{cct}.

Again, when Coulomb attraction and driving field amplitude become
comparable, the classical dynamics of the Rydberg electron
turns chaotic and phase space is divided into
regions of regular and irregular motion. At sufficiently large field
amplitudes, only the principal resonance between the driving frequency
and the Kepler motion is left as an (elliptic) island of regular motion in the
chaotic sea \cite{ab1}. Unlike the CPM case, the stable periodic orbit at the
center of the elliptic island is not characterized by a set of
simple analytical expressions. However, the local oscillatory motion
can be plugged into the form of a Mathieu equation \cite{bz},
 the numerical
solution of which provides good estimates of the energy of the quantum
mechanical ground state and of the first excited states
of the local hamiltonian \cite{holt,sk}.
The oscillator ground state is the
wave packet eigenstate of the atom in the field and
follows the classical, periodic evolution of the principal resonance
island \cite{hh,ab1}. Depending on the value of the classical angular
momentum and
its projection on the symmetry axis, the wave packet may probe
the Coulomb singularity and consequently displays some transient
dispersion which mimicks the acceleration of the classical particle at
the aphelion \cite{ab1}.

Since, in the LPM case, the 
numerical detection of the wave packet is less
straightforward than for CPM, we restrict ourselves to the
investigation of the hydrogen atom confined to one spatial dimension,
along the field polarization axis. As the driving field amplitude is
increased, it is this direction along which chaos is born in the full
dynamics of the 3D atom \cite{ab1,i3e,ab2},
and therefore this approximation will be
sufficient for our present purposes.

The LPM  case has the same scaling property than the CPM case:
the classical dynamics depends only on the scaled field
$F_0=F\omega^{-4/3}$.
Numerically, the wave packets are
identified by their
large (as compared to other nearby Floquet states)
overlap with the state $\mid n_0\rangle$ \cite{comment}, which is
resonantly coupled to its nearest neighbour $\mid n_0+1\rangle$ by the
driving frequency.
Similarly to the CPM case, the quasienergies $E_i$ and
resonance widths $\Gamma_i$ of the wave packets
are used for the statistical analysis.
All data samples are characterized by a fixed value of the
parameter $F_0$, hence they correspond to the same
 structure of classical phase space. For a given value of
$F_0$, $\omega$ (and, accordingly, $F$) has been scanned to give
$1000$ eigenvalues per $F_0$ value.

\subsection{Numerical results}

The typical behaviour of the wave packet ionization width versus
the microwave frequency has already been presented  in \cite{zdb95}
for a fixed microwave amplitude. It displays very strong fluctuations
over several orders of magnitude for small changes of the frequency
(typically of the order of 1 part in 1000).
These fluctuations -- although perfectly deterministic --
look completely random and are strongly reminiscent of the
universal conductance fluctuations observed in mesoscopic systems \cite{meso}.
Indeed, the ionization width measures the rate at which an electron
initially localized close to the stable resonant trajectory ionizes,
i.e. escapes to infinity. In other words, the ionization width
directly measures the 
conductance of the atomic system from the initial
point to infinity.
In the quantum language, the ionization width is due to the coupling
(via tunnel effect) between the localized wave packet and states
lying in the chaotic sea surrounding it. While the energy (or quasi-energy
in the LPM case) of the wave packet is a smooth function of the
parameters $F$ and $\omega,$ (see Eq.~(\ref{semicl})), the energies
of the chaotic states  display a complicated behaviour characterized
by level repulsion and strong avoided crossings. By chance, it may happen
that -- for specific values of the parameters -- there is a quasi-degeneracy
between the wave packet eigenstate and a chaotic state. There, the two
states are more efficiently coupled by tunneling and the ionization
width of the wave packet eigenstate increases. 
This is the very origin
of the observed fluctuations. Simultaneously, the repulsion between
the two states should slightly modify the energy (real part of the complex
eigenvalue) of the wave packet state. A simple way of measuring this
effect is to compute also the shift of the real part of the
energy level with respect to its 
semiclassical position (which does not exhibit the repulsion from a 
near-degenerate state).

As mentioned above, we study the
fluctuations for a fixed value of the classically
scaling parameter $F_0$ versus $n_0=\omega^{-1/3}.$
Exemplary ionization
width and {\it level shift} fluctuations are presented in Fig.~\ref{fig1}
on a logarithmic scale for the CPM case.
Note that both quantities fluctuate over
several orders of magnitude and that the widths are more sensitive
to changes of $n_0$. Since the shifts can take both positive
and negative values, the absolute value of shifts is plotted.

Importantly we must mention that the shifts plotted in Fig.~\ref{fig1}
and used later for the statistical analysis are not obtained directly
from the difference between the resonance energy and the semiclassical
prediction, Eq.~(\ref{semicl}), as anticipated before. 
These differences show
a bias -- the average shift is non zero. It indicates that although
Eq.~(\ref{semicl}) well predicts the wave packet energy (to within a
fraction of the mean level spacing), the remaining difference is
not solely due to the fluctuations. There is a slowly varying
part in it which most probably originates from the unharmonic
corrections. The latter could be estimated; however, since in the LPM case,
we do not have any good semiclassical prediction, we find the fluctuating
part of the shift in both cases by 
subtracting from the exact quantum
energies the smooth background. The
latter is obtained by a low order polynomial fit of the wave packet
eigenenergies as a function of the parameter $n_0.$ 

To describe the fluctuations {\it quantitatively}, which is the main
aim of this study, we calculate the statistical distributions
of the ionization widths $P(w)$ and of the energy shifts
$P(s).$ Typical distributions are displayed as histograms
on a double logarithmic scale in Fig.~\ref{fig2}.
The data are those of Fig.~\ref{fig1}.
The use of logarithmic scales is useful to show quantitatively
the fluctuations over several orders of magnitudes.
>From this figure, we immediately obtain the following qualitative conclusions:
\begin{itemize}
\item
The distribution of energy shifts $P(s)$ tends to a constant as
$s \rightarrow 0.$
\item Above some critical value, the distribution drops and decreases
roughly algebraically with $s.$ The slope is close to $-2$, indicating
a $P(s) \equiv 1/s^2$ behaviour.
\item Finally, $P(s)$ falls off abruptly above some value.
\item The distribution of ionization widths $P(w)$ behaves algebraically
with $w$ at small width, with a slope close to $-1/2.$
Hence, the small widths are the most probable ones.
\item Above some critical value, the distribution drops faster,
roughly with a $1/w^{3/2}$ behaviour.
\item Finally, similarly to the shift distribution, there is a
final sharp cutoff.
\end{itemize}

These conclusions are similar to the ones obtained in a slightly
different physical situation, when two symmetric regular islands are
coupled to a single chaotic sea (see Ref.~\cite{TU94} for a complete
discussion of this physical situation). There, states lying on the
regular islands appear in doublets with even/odd parities with respect
to the discrete symmetry exchanging the two regular islands. The splitting
of the doublet is a direct measure of the chaos assisted process where
the particle tunnels from one regular island, then diffuses in the chaotic sea
and finally tunnels to the other regular island. This process is very similar
to the one studied in the present paper where the particle tunnels and
then diffuses towards infinity.
The splitting distribution observed in \cite{TU94,LU96} is indeed very
similar to our shift distribution as explained below.

In the following section, we propose a simple model - mainly based on 
physical ideas similar to those used in \cite{TU94} - which allows
us to understand
the properties of our shift and width distributions.

\section{Statistical Model}

We shall consider a simple statistical model amenable to an
analytical treatment. Despite its simplicity, we shall see
that it is capable of
describing quantitatively the fluctuations of resonance widths and
shifts observed in the numerical analysis of the microwave ionization of
hydrogen atoms.

The model system is directly based on the understanding
we have of the origin of the fluctuations and follows the idea
already used in~\cite{TU94}.
The idea is to consider the wave packet eigenstate as coupled
randomly to a set of chaotic states (described by Random Matrix Theory)
which are 
themselves randomly coupled to the atomic continuum.
More precisely, our model consists of a single state $|0\rangle$ (representing a
state localized on the elliptic island) coupled (weakly) to
the space of $N$ ``chaotic'', irregular levels, which are close to 
$|0\rangle$ in energy.
The latter
states are considered to be, strictly speaking, resonances
rather than bound states due to the mechanism responsible
for decay (e.g., the ionization of an atom).
While we consider
a single localized state in the model, there may be several other
states localized on the same island. They will have typically
much different energies and their mutual coupling via the
chaotic states is negligible.

The Hamiltonian
matrix may be represented as a $(N+1)\times (N+1)$ matrix:
\begin{equation}
{\cal H}=\left(
\begin{array}{cc}
0 & \sigma \langle V| \\
\sigma |V\rangle & H_0-i {\cal W}
\end{array}
\right),
\label{mat1}
\end{equation}
where we have assumed that the energy of the localized state sets
the zero of
the energy scale.
The first vector in the basis represents the localized state,
the following $N$ ones the chaotic states.
In the chaotic subspace, the statistical properties of the Hamiltonian
are well represented by a $N\times N$ random matrix $H_0$.
We shall deal with
problems with a preserved (generalized) time-reversal invariance --
$H_0$ should belong, therefore, to
the Gaussian Orthogonal Ensemble (GOE) of real symmetric matrices~\cite{haake,bohigas}.
The matrix elements of $H_0$ are independent random Gaussian variables:
\begin{equation}
P((H_0)_{ij}) = \frac{\exp \left(-\frac{(H_0)_{ij}^2}{2\sigma_{ij}^2} \right)}
{\sqrt{2\pi \sigma_{ij}^2}},
\end{equation}
with
the variance satisfying
\begin{equation}
\sigma_{ij}^2 = (1+\delta_{ij}) \frac{\pi^2 \Delta^2}{N}.
\end{equation}
Here, $\Delta $ is the mean level spacing between consecutive
chaotic states close to energy 0.

Following a commonly accepted approach \cite{HILSS}
the coupling of the chaotic state
to the continuum is introduced by the decay matrix:
\begin{equation}
{\cal W} = \frac{\gamma}{2} |W\rangle \langle W|
\label{w}
\end{equation}
where the $N$ component real vector $|W\rangle $ describes the coupling of
the chaotic states to the continuum.
As in Ref.~\cite{HILSS}, we take this
vector to be composed of Gaussian
distributed random numbers with vanishing mean and unit variance.
The real coefficient $\gamma$ measures the strength of the
decay to the continuum.
Such a form implies that there is only one significant
decay channel for chaotic states. This is far from obvious and,
as discussed below, probably true only at relatively low field
strength.
When there are several open ionization channels, a convenient form
of the decay matrix is~\cite{HILSS}:
\begin{equation}
{\cal W}=\sum_{k=1}^{M} \frac{\gamma_k}{2} \ |W^{[k]}\rangle \langle W^{[k]}|,
\label{Gam}
\end{equation}
where $M$ denotes the number of open channels
(degeneracy of the continuum) and the $N$ component
real
vectors $|W^{[k]}\rangle $ describe the coupling of the chaotic states
to channels $k=1,..,M$. Again, we take these
vectors to be composed of Gaussian
distributed random numbers with vanishing mean and unit variance.
The $\gamma_k$ real coefficients measure the strength of the
decay to continuum $k$.

In a similar way, the (real) $N$-component vector $|V\rangle $ in
Eq.~(\ref{mat1}) describes the
coupling of the localized state $|0\rangle$ to the chaotic states.
Each component of $|V\rangle $ is taken as a Gaussian distributed random
number of zero mean and unit variance. The coefficient $\sigma$ in
Eq.~(\ref{mat1}) is a
measure of the strength of the coupling between the localized state
and the chaotic subspace.

If the coupling to the continuum is neglected $(\gamma=0)$, the model
describes a single bound state randomly coupled to $N$ chaotic states.
This is exactly the model succesfully used in~\cite{TU94} to describe the 
splitting of doublets induced by chaos assisted tunneling.

The model has several free parameters: the mean level
spacing between chaotic states, $\Delta$, the strength of the
coupling with the ionization channel, $\gamma$,
and the strength of the coupling to the localized state, $\sigma$.
There is a trivial scaling law of the Hamiltonian ${\cal H}$, Eq.~(\ref{mat1}),
which implies that, except for a global multiplicative factor, there
are only two relevant dimensionless parameters,
$\sigma/\Delta$ and $\gamma/\Delta$ in the model.
For several open channels, the relevant parameters are $\sigma/\Delta$ and
$\gamma_k/\Delta,\ k=1..M.$

Due to the interaction with ``chaotic" resonances, the state $|0\rangle$
is not an eigenstate of the full Hamiltonian ${\cal H}$, Eq.~(\ref{mat1}).
However, in most cases, the coupling to chaotic resonances is weak
and the true eigenstate does not differ very much from $|0\rangle :$ this
is the perturbative regime that we shall consider in the rest of this paper.
This regime is obtained when the coupling is much smaller than the mean
level spacing between chaotic states, i.e.:
\begin{equation}
\sigma \ll \Delta .
\label{pert1}
\end{equation}

We will see below that this condition is always satisfied for the
real physical system (hydrogen atom + microwave field) for the microwave
fields studied in this paper. At very high field, when the nondispersive
wave packet is destroyed, this pertubative approximation should break down.
The physical interpretation is clear: if Eq.~(\ref{pert1}) is not satisfied,
the localized state is spread over several eigenstates of ${\cal H}$ and
completely looses its identity. It has been shown elsewhere \cite{zbd96} that,
in the perturbative regime, the localized state can be interpreted as a soliton
weakly interacting with the background of chaotic states, but
essentially keeping its shape when a parameter (for example the microwave
field strength) is varied.

In the following, we will also assume that the ionization rates (widths)
of the chaotic states are small compared to their mean level spacing,
i.e. that the decay matrix $W$, Eq.~(\ref{Gam}), can be considered as
a pertubation. This implies:
\begin{equation}
\gamma \ll \Delta .
\label{pert2}
\end{equation}
Such a condition is not {\em strictly} necessary, but it makes calculations
simpler. Physically, it means that the various chaotic
resonances are isolated. This is a typical situation in our system -
the ionization of an atom in
the presence of microwave driving for not too strong microwave amplitudes.

With the above assumptions -- motivated by the physics of the process
studied --
the shift and width of the localized state may be obtained
using the lowest nonvanishing order of perturbation theory.
Such an approach is justified unless an accidental degeneracy
between the localized state and one of the eigenstates of the matrix
$H_0$ occurs. Neglecting such degeneracies (which only affect the tail of
the distribution, see Sec.~IV.A below) and performing
an average over the random matrix ensemble defined by Eq.~(\ref{mat1})
makes it possible to extract
the analytic expressions both for the distribution of shifts
and for the distribution of widths. The details
of the derivation are given in the Appendix.
We give here only the important results.

The shift distribution  is obtained
along a similar way to Leyvraz and Ullmo
\cite{LU96} and takes the form of a Cauchy law
\begin{equation}
P(s)=\frac{1}{\pi}\frac{s_0}{s_0^2+s^2},
\label{eqs}
\end{equation}
with
\begin{equation}
s_0=\frac{\pi \sigma^2}{\Delta}.
\label{s0ab}
\end{equation}
Importantly, this result is independent of the
degree of correlation among the eigenvalues of the $H_0$ matrix: the
same result is obtained for an uncorrelated spectrum (Poisson-like
distributed eigenvalues, physically corresponding to coupling
of the localized state with a set of ``regular" states, instead of
chaotic ones as in the model described above),
a GOE or a picket fence (harmonic ocsillator)
spectrum \cite{LU96}. This Cauchy distribution is the same than the one
obtained \cite{TU94,LU96} in the absence of ionization.

The situation is a bit more complicated for the width distribution.
For Poissonian distributed
eigenvalues of $H_0$ (uncorrelated spectrum), a similar Cauchy distribution
is obtained for the {\em square root}
of the width,
corresponding to the
following distribution of widths (see Appendix):
\begin{equation}
P_{\rm Poisson}(w)=\frac{1}{\pi}\frac{\sqrt{w_0}}{\sqrt{w}(w+w_0)},
\label{eqx}
\end{equation}
with
\begin{equation}
w_0=\frac{4\sigma^2\gamma}{\Delta^2}.
\end{equation}
For a matrix $H_0$ belonging to the GOE, the distribution
is slightly more complicated (see Appendix):
\begin{equation}
P_{\rm GOE}(w) = \frac{2}{\pi^2} \frac{{w'_0}^{3/2} \ln (w+\sqrt{1+w/w'_0})
+ \sqrt{ww'_0(w+w'_0)})}{w(w+w'_0)^{3/2}}
\label{eqx2}
\end{equation}
with
\begin{equation}
w'_0= \frac{\pi^2 \sigma^2 \gamma}{\Delta^2} = \frac{\pi^2}{4}\ w_0.
\end{equation}
Although it is distinct from Eq.~(\ref{eqx}), it has in fact quite a similar
shape, see Fig.~\ref{fig2.2}.
In particular, if $P_{\rm GOE}$ is rescaled by a global multiplicative
factor of about 20\%, it is almost indistinguishable from $P_{\rm Poisson}.$
Quite a large statistics is necessary to determine which of
the two distributions is a correct one for a given data set (see
also next Section).

We can also obtain the distribution of widths for the chaotic states in
the perturbative regime. This is the well-known ``non-overlaping resonances"
regime~\cite{brody,bg:prl,dupret} where the widths are distributed according
to a Porter-Thomas distribution:
\begin{equation}
P_{PT}(w)=\frac{1}{\sqrt{w\overline{w}}}
\exp(-w/2\overline{w}),
\label{PT}
\end{equation}
where $\overline{w}$ denotes the average width.
For small $w$, $P_{PT}$ diverges as $w^{-1/2}$ while
for large $w$ it decays exponentially.

\section{Analysis of the data}

\subsection{Quantitative analysis of the fluctuations with the statistical
model}

The exact expressions, Eqs.(\ref{eqs}-\ref{eqx2}), obtained
in the perturbative regime, reproduce qualitatively most of the
statistical distributions numerically observed for
the shift and width of the nondispersive wave packet
of the hydrogen atom in a microwave field, see Fig.~\ref{fig2}.
For the shift distribution $P(s),$ the distribution is constant
near $0$ and decays like $1/s^2$ at large $s$. The only difference
is the absence of the sharp cut-off in the perturbative expression.
This can be easily understood: the large energy shifts correspond
to quasi-degeneracies between the localized state and one specific
chaotic state, i.e. to the immediate vicinity of an avoided crossing.
There, the simple pertubative scheme breaks down and the actual shift
remains finite as the perturbative expression diverges as the inverse
of the unperturbed spacing between the chaotic and localized states
Hence,
the actual distribution has to fall faster than the perturbative one
at large shifts, as numerically observed.

The width (ionization rate) distribution behaves similarly.
Both the initial $1/w^{1/2}$ regime and the following $1/w^{3/2}$ regime
observed for the 
wave packet
are well reproduced by the simple statistical model. Again, the difference
is the absence of the cut-off for very large widths. 
The reason is identical,
namely the breakdown of the perturbative approximation.

Nevertheless, we can go beyond the perturbative scheme, using
the statistical model described in the previous section, but calculating
numerically the shift and width distribution.
This has been done by numerical diagonalization of the complex Hamiltonian,
i.e. of
random matrices
corresponding to Eq.~(\ref{mat1}), generated according to the rules
given above. A diagonalization of a matrix of size
$N=80$ yields a single shift and width for chosen values of
$\sigma/\Delta$ and $\gamma/\Delta$.
Using different 
random matrix realizations
we accumulate up to 50~000 data for comparison. We have verified
that the distributions obtained do not depend on the matrix size $N$.

In Fig.~\ref{fig2.1}, we show the numerical results
for the shift distribution obtained for 
the hydrogen atom in a circularly polarized
microwave field with the
perturbative analytical expression for our random matrix model
(pure Cauchy distribution),
Eq.~(\ref{eqs}), and with the
full non-perturbative result using our statistical model,
both on a linear scale 
-- well suited for small and moderate
shifts -- and on a double logarithmic scale -- well suited for the
tail of the distribution at large shifts.
As expected, the perturbative analytical expression reproduces
the numerically observed distribution, except for the exponentially
small tail at large shift. The full non-perturbative distribution
is found to be in excellent agreement with the numerical data
for the real system -- the hydrogen atom in a circularly polarized
microwave field, which proves that our simple statistical model
actually catches the physics of the chaos assisted tunneling
phenomenon. A similar conclusion has been reached \cite{TU94} for
doublet splittings induced by chaos assisted tunneling.

In Fig.~\ref{fig2.2}, a similar analysis is done for the
distribution of widths, both on linear and
double logarithmic scale. On the linear scale, instead
of the width distribution,
Eq.~(\ref{eqx2}), itself which diverges at zero
(see Appendix), we plotted
the distribution for the square root of the width,
which tends to a constant value at zero. As can be seen,
the agreement is again excellent over the full range,
the perturbative expression being inaccurate in the tail only,
as expected. In addition to the
perturbative analytical expression, Eq.(\ref{eqx2}),
we have also drawn the distribution
expected when the states in the chaotic sea have eigenergies
described by a Poisson distribution rather than a GOE one, Eq.~(\ref{eqx}).
Both distributions are similar far from the origin and differ by
about 20\% at $w=0.$ At first glance, it seems that the Poisson
curve agrees slightly better than the GOE curve, which is somewhat
surprising and not understood
as chaotic motion surrounding the stable island
suggests the choice of the GOE. However, the deviation is at the border
of statistical significance.
In the double-logarithmic plot, we have also added the Porter-Thomas
distribution, Eq.~(\ref{PT}), which reproduces correctly the tail
at large widths.

It is remarkable that both the shift and the width
distributions are so well reproduced by the random matrix
model.
This proves that our simple statistical model carries the essential part
of the physics. The data presented are the first, as far as we know, manifestation
of the chaos assisted tunneling process in a realistic, experimentally
accessible systems.

Even if the perturbative expression, Eq.~(\ref{eqx}),
incorrectly describes the
non-perturbative large width tail,
it is
clear that the initial $w^{-1/2}$ decrease, similar to
$P_{PT}$, Eq.~(\ref{PT}), is followed by a regime of $w^{-3/2}$ behaviour.
The length of this $w^{-3/2}$ behaviour provides an interesting
information about the system. For strong coupling, $\sigma \ge \Delta,$
the localized state $|0\rangle$ is strongly mixed with the chaotic state.
Thus, its width
distribution would be the same as that of other resonance states, i.e.
a Porter-Thomas distribution. Hence, the $w^{-3/2}$ part
there shrinks to zero and the power $-1/2$ law is followed immediately by
the exponential tail.
The relative importance of the $w^{-3/2}$ part in the
width distribution indicates, therefore, the presence of the
weak coupling, perturbative regime.
Compared to the pure chaotic state where fluctuations
of the widths are already known to be large, the effect of additional tunneling
is to shift some part of the width distribution towards small
widths while keeping the exponential cut-off, that is
to increase the fluctuations. In the perturbative limit, the fluctuations
become so large that the average width 
{\em diverges}.

The success of the statistical model allows to give a complete
physical interpretation of the observed data:
\begin{itemize}
\item The smallest shifts and widths, observed for
\begin{eqnarray}
s < s_0\\
w < w_0
\end{eqnarray}
with probabilities behaving, respectively, as $s^0$ 
and $w^{-1/2}$,
correspond to the localized state lying far from quasi-degeneracies
with one of the chaotic states. Then, the localized state is weakly
contaminated by the various surrouding chaotic levels. For example, its shift
is the sum of the effect of level repulsion by the various chaotic states.
Chaotic states with higher (resp. lower) energy push the localized state down
(resp. up) in energy, which globally results in a small shift with a random 
sign.
The width results from the interference between the elementary ionization 
amplitudes
contributed by 
the various chaotic states. 
As there is only one open decay channel, the
amplitudes -- not the probabilities -- have to be added. 
As they are essentially random
uncorrelated variables, the interference is mainly destructive, producing
small widths with a large probability.

\item The intermediate shifts and widths, observed for
\begin{eqnarray}
s_0 < s << \sigma\\
w_0 < w << \gamma
\end{eqnarray}
with probabilities behaving respectively as $s^{-2}$ and $w^{-3/2}$,
correspond to one chaotic state being much closer in energy
to the localized state than to the other chaotic states, but
nevertheless sufficiently far to be 
coupled only perturbatively. Then,
the single coupling to this particular chaotic state dominates both
the shift and the width of the localized state.

\item The largest shifts and widths, observed for
\begin{eqnarray}
s \simeq \sigma \\
w \simeq \gamma
\end{eqnarray}
correspond to the quasi-degeneracy between the localized state and one chaotic
state, with strong non-perturbative mixing between the two.
This is where the exponentially decreasing tails are observed.
\end{itemize}

In the latter two cases, as one chaotic state is dominant, approximate
expressions for the distributions can be obtained by a simple
two-level model. Although the expression of the shift and the width are
then easy to obtain (diagonalization of a $2\times 2$ complex matrix),
we have not been able to perform analytically the averaging
over the ensemble of random matrices.

\subsection{Extraction of the relevant physical parameters from the statistical data}

The simplest possible measures of the fluctuations of a quantity are typically
its average value and the variance. Inspection of
the perturbative distributions for both the shifts, Eq.~(\ref{eqs}), and
the widths, Eq.~(\ref{eqx}), suggests, however, some caution.
Indeed, the average value of the width is not defined (the corresponding
integral diverges) and the same is true for the variance of the shift
(the average value is zero due to the symmetry of the distribution).
This is because of the existence of long algebraic tails, $1/s^2$ and
$1/w^{3/2}$, in the distributions. The variance of the shift and the
average value of the width are infinite because of the diverging contributions
of these tails. This is an example of unusual random processes such as Levy
flights~\cite{levy} where rare events play the dominant role.
Ultimately,
it is because, in perturbation theory, the contamination of a state by its
neighbors is proportional to the ratio of the coupling matrix element
to the energy difference and consequently decreases slowly: even a far distant
level can have a large effect.

For the full non-perturbative 
distributions, the exponential cutoff
at large values kills the divergence of the various integrals and
average values and variances -- as well as higher moments of the
distributions -- are well defined and could be calculated.
Their precise values, however, depend crucially on the position of the cutoffs.
Hence, distributions identical for small widths (shifts) and only
differing at large widths (shifts) may have completely different
average values and variances.
Calculating such quantities requires a very accurate knowledge
of the tails of the distributions. The average values and the
variances are thus fragile and difficult to calculate on a real
system like the hydrogen atom in a microwave field; they do not provide
us with the most interesting physical information.

Rather than the average values, we prefer to define typical values.
The typical width $\tilde{w}$ lies at the middle of the distribution, 
such that half of the widths are larger and half of them smaller, i.e.
\begin{equation}
\int_0^{\tilde{w}} P(w)\ dw = \frac{1}{2}
\end{equation}
With such a definition, the typical width is robust, only weakly sensitive
to the tail of the distribution.

A similar definition holds for the typical shift, slightly modified
to take into account that $s$ can be either positive or negative:
\begin{equation}
\int_0^{\tilde{s}} P(|s|)\ d|s| = \frac{1}{2}.
\end{equation}

For the perturbative distributions in the statistical random matrix model,
the typical widths and shifts can be calculated from
Eqs.~(\ref{eqs}),(\ref{eqx}),(\ref{eqx2}):
\begin{equation}
\left\{
\begin{array}{lll}
\displaystyle \tilde{s}&=&\displaystyle  \frac{\pi \sigma^2}{\Delta}\\
\displaystyle \tilde{w}&=& \displaystyle \frac{4 \sigma^2 \gamma}{\Delta^2}\ \ \ \ {\rm for\ a\ Poisson
\ spectrum}\\
\displaystyle \tilde{w}&=& \displaystyle \frac{5.48 \sigma^2 \gamma}{\Delta^2}\ \ \ \ {\rm for\ a\ GOE
\ spectrum}.
\end{array}
\right.
\label{typical}
\end{equation}
For the full non-perturbative 
random model distributions -- as long as the basic hypothesis
of small coupling, Eqs.~(\ref{pert1}) and (\ref{pert2}), are true --
we carefully checked that
the typical widths and shifts are not significantly 
(within few tenths of percent)
different from the previous analytic expressions.

For a real physical system like the wave packets 
in the
hydrogen atom exposed to a microwave
field, we can reliably extract from the statistical data the typical
width and shift. We have also compared the numerically obtained
distributions with non-perturbative distributions of our statistical
model and performed best fits of the former by the latter. Again, we checked
that the typical width and shift for the best fit do not differ
by more than 
few tenths of percent from the direct measures. This implies
that the typical shift and width can be safely used to extract
from the statistical distributions the relevant 
physical parameters
$\sigma$ (coupling between the localized
state and the chaotic states) and $\gamma$ (ionization widths of the
chaotic states). Slightly different values are obtained
if the Poisson or GOE expressions are used.
In the following, we have used the GOE expression.

In fact, only $\sigma^2/\Delta$ and
the dimensionless parameter $\gamma/\Delta$
(from the ratio $\tilde{w}/\tilde{s}$) can be easily extracted.
Obtaining the other dimensionless parameter $\sigma/\Delta$ or
$\sigma$ and $\gamma$ themselves requires to know the mean spacing $\Delta$ 
between
chaotic resonances.
Surprisingly, this is not straightforward.
To understand the problem, consider
the diagonalization of the Floquet Hamiltonian for the LPM case. The
number of states present in a single Floquet zone depends on the
number of photon blocks included in the diagonalization. When
it is increased, new states appear in the vicinity of the wave packet state
corresponding to either low lying atomic states (with very different
energy but shifted upwards by an integer times the photon frequency) or
highly excited states or resonances (shifted downwards).
These states should not 
contribute to the determination
of $\Delta$ since they have a vanishing overlap with atomic states
building the wave packet.
Hence, the mean level spacing $\Delta$ between 
chaotic states
is a somewhat ambiguous quantity. However -- as will be seen 
at the beginning of the next section --
all our results are obtained either in the perturbative regime
$\sigma ,\gamma \ll \Delta$ or close to it, and the mean spacing
$\Delta$ is just a scaling parameter. A rough estimate of $\Delta$
is obtained assuming that only states with similar principal quantum number,
let us say differing by less than a factor 2, are efficiently coupled.
Experimental results on hydrogen atoms in a linearly polarized microwave field
\cite{koch}
suggest that the physics of ionization is entirely dominated
by states with principal quantum number less than $2n_0$
(when the experimental cut-off is changed from infinity to $2n_0$,
no important change is observed). At low microwave field,
the number of efficiently coupled states is smaller,
but this is not the regime of chaotic diffusion we are interested in.
Chaotic motion requires overlap between classical resonance islands,
i.e. efficient coupling between states of largely different principal
quantum numbers.
This gives the following approximate mean level spacing, used later in
this paper:

\begin{equation}
\Delta\approx \frac{1}{n_0^4}.
\end{equation}

\subsection{Tunneling and chaotic ionization rates}
\label{sectyp}

Fig.~\ref{fig3} shows the typical width and shift for the non-spreading
wave packet of the hydrogen atom in a circularly polarized microwave field
at frequency $\omega=1/(40)^3,$
as a function of the scaled electric field $F_0=F(40)^4.$
Each point in this curve results from the numerical diagonalization
of several hundred matrices, each of typical size several tens
of thousands, for neighbouring values of the microwave
field strength and frequency.
The statistical model described above makes it possible to separate
the intrinsic huge fluctuations of the ionization rate
and extract values of the various couplings.
This is very clear in Fig.~\ref{fig3} where both the typical width and the
typical shift are relatively smooth functions of the field strength,
with short range fluctuations smaller than a factor 2, whereas the
raw shift and width display fluctuations over 
at least 3 orders of magnitude,
compare Fig.~\ref{fig3} with Fig.~\ref{fig1}.

In Fig.~\ref{fig3}, one can easily check that the typical width
is always smaller than the typical shift by at least 
one order
of magnitude. As the ratio of the two is $\gamma/\Delta$, see
Eq.~(\ref{typical}), this implies that inequality (\ref{pert2})
is verified.
Also, the typical shift is smaller than the mean level spacing
(represented by the dashed line) which, using 
Eq.~(\ref{typical}),
shows that inequality (\ref{pert1}) is also verified.
Altogether, this proves that our data are effectively obtained
in the perturbative regime.

The third observation in Fig.~\ref{fig3} is that neither the typical
width, nor the typical shift are monotously increasing functions of the
microwave field strength, but display various bumps. These bumps
are obviously strongly correlated, which indicates that they
are due to variations of the tunneling rate $\sigma$ rather than
variations of $\gamma.$
Indeed, in Fig.~\ref{fig4},
we plot the dimensionless parameters $\sigma/\Delta$ and $\gamma/\Delta$
deduced from the typical shift and width.
It confirms that the tunneling rate $\sigma/\Delta$ is a slowly increasing
function of the
field strength with various structures. The bumps occur
precisely at values of the field strength where there is
a resonance between the
eigenfrequencies $\omega_+$ and $\omega_-$, see Eq.~(\ref{modes}),
of the motion in the vicinity of the stable fixed point supporting
the non spreading wave packet. This has been analyzed in reference~\cite{zbd96}
where it is shown that the bump around $F_0=0.023$ corresponds to the
1:4 resonance and the bump just below $F_0=0.04$ to the 1:3 resonance.
In the vicinity of such a resonance, the classical dynamics is strongly
perturbed, some secondary resonant tori and islands appear. The bumps
in Fig.~\ref{fig4} are just quantum manifestations of an increased
transport rate induced by these classical resonances.
Not surprisingly, $\gamma/\Delta$, which represents the ionization rate
of states surrounding the resonance island, is practically
not affected by these resonances (only small residual
oscillations are visible around $F_0=0.04$). On the other hand, it increases
very fast up to scaled field $F_0\approx 0.04$
where it saturates to a roughly constant value. This has a simple
semiclassical explanation. Below $F_0\approx 0.04$, chaos is not established
around the principal resonance island, there still exist some regular
tori further in phase space which strongly slow down the classical
chaotic diffusion. Above 0.04, only the principal resonance island survives,
the chaotic ionization rate is quite large ($\gamma/\Delta$ is of the order
of 0.1) and only slowly increases with the
field strength.

Strictly similar observations can be made for the hydrogen atom exposed
to a linearly polarized microwave field, which proves that they
are not specific to one system under study, but rather general
properties of chaos assisted tunneling followed by chaotic
diffusion. Fig.~\ref{fig5} displays the typical width and typical shift
of the non-spreading wave packet for $n_0=40$, as a function
of the scaled microwave field $F_0=Fn_0^4.$ Again, these data are
obtained in the perturbative regime, see Eqs.~(\ref{pert1})-(\ref{pert2}),
and display obviously correlated bumps. The dimensionless tunneling
rate $\sigma/\Delta$ and chaotic ionization rate $\gamma/\Delta,$
shown in Fig.~\ref{fig6}, indicate that the bumps are due to
secondary resonances
inside the primary resonance island between the external microwave frequency
and the internal Kepler motion. Comparison between Fig.~\ref{fig4}
and Fig.~\ref{fig6} shows that both the tunneling rate
 and chaotic ionization rate
are of the same order of magnitude in linear and circular polarization,
with similar changes versus $F_0$, up to possibly a roughly
constant multiplicative factor.
This is a confirmation of the experimental observation that
very similar ionization threshold frequency dependences
 are observed in the two cases provided $F_0$ is appropriately
rescaled
\cite{koch:circ}. As in \cite{koch:circ} we observe that larger
values of $F_0$ are necessary in LPM to result in the similar
behaviour as for CPM.

To make the study complete, we have also studied how the
typical width and the typical shift change when the principal
quantum number $n_0$ -- or equivalently, the microwave
frequency $\omega=1/n_0^3$ -- is changed.
The result is shown in Fig.~\ref{fig7} for the circular polarization,
for a fixed scaled microwave field $F_0=0.0426.$ In this plot,
the classical dynamics of the system is absolutely fixed, the only
varying parameter being the effective Planck's constant
$\hbar_{\rm eff}=1/n_0.$ The striking phenomenon is the fast decrease
of both the typical width and typical shift with $n_0.$ In the logarithmic
scale of Fig.~\ref{fig7}, is appears as a straight line indicating
an exponential decrease with $n_0.$ Also, the two quantities
decrease along parallel lines, which -- according to 
Eq.~(\ref{typical}) --
indicates that the tunneling rate $\sigma$ is responsible
for this decrease.
In Fig.~\ref{fig8}, we plotted the dimensionless tunneling
rate $\sigma/\Delta$ and chaotic ionization rate $\gamma/\Delta$ as
a function of $n_0.$ Note that $\sigma/\Delta$ is plotted using a logarithmic
scale and $\gamma/\Delta$ on a linear scale!
The exponential decrease is of the form:
\begin{equation}
\sigma/\Delta \simeq \exp (-n_0S) = \exp \left( - \frac{S}{\hbar_{\rm
eff}}\right)
\end{equation}
with $S\approx 0.06 \pm 0.01$
(extracted from the plot).

Such a dependence is typical for a tunneling process. $S$ then
represents the tunneling action from the stable fixed point
where the wave packet sits to the chaotic region surrounding the
resonance island. If complex orbits are used, 
$S$ can be thought
as the imaginary part of the action of a complex tunneling
orbit \cite{whelan,PROF.DR.AMAURY-MOUCHET}. In our realistic system, finding such complex orbits
is much more difficult than in the few
model systems where this analysis has been done. We have not been able to find
the complex path associated with the tunneling process, but
our purely quantum results may provide a guide in this search, as
they show that the imaginary action has to be of the order of 0.06
for $F_0=0.0426.$

On a linear scale, see Fig.~\ref{fig9}, the dimensionless chaotic ionization
rate
$\gamma/\Delta$ is a slowly increasing function of $n_0.$ A simple
classical analysis using the so-called Kepler map \cite{i3e}, which is known
to produce relatively good predictions for the ionization
threshold of Rydberg states by a microwave field \cite{zgd96,bd93}
predicts a linear dependence versus $n_0,$ while the numerical result
seems rather a quadratic function. 
This discrepancy
could be due either
to the approximations done to obtain the Kepler map
or to the fact that, for high $n_0,$ the statistical model
used to extract $\gamma/\Delta,$ see 
Eqs.~(\ref{eqs}),(\ref{eqx}),(\ref{eqx2}),(\ref{typical}), is no
longer valid because several ionization channels are open (see next section).

Let us note that to get 800 data points
for $n_0=100$ (enough to determine the typical shift
and width) requires about 40 hours of Cray J98 single
processor CPU time. The presented results are, in this sense,
quite costly (the size of diagonalized matrices exceeded 200 000
in this case).

\subsection{Limitations of the model}

Although the simple statistical model well describes the fluctuations
of the width and shift of the non-spreading wave packet
for a range of scaled microwave field ($F_0\in [0.04,0.08]$ for
LPM and $F_0\in [0.02,0.06]$ for CPM, both for $n_0$ around 40) --
the region
where experiments are usually done -- additional difficulties
appear for lower and higher field values.

For lower $F_0$ values, a statistically
significant part of the data show very small widths, at the limits
 of the numerical precision. 
At the same time, 
plots of the wave packets similar to those shown 
 in \cite{dzb95} suggest that the states are more extended,
extending far from the stable island. The 
situation then may not correspond
to a clear-cut case of the chaos assisted tunneling process. 
Our LPM data 
indicate 
that in such cases the singularity 
for small widths is much
stronger than $\Gamma^{-1/2}$.
Similarly, in the CPM case,
we did not present the random matrix fit for $F_0=0.038,$ as a
significant part of the data is 
affected there by a strong classical
$1:3$ resonance \cite{zbd96}. Thus we do not 
face a clear case
of a single localized state  but
rather two strongly coupled localized states
decaying via a chaos assisted tunneling process. Since such a case
is quite rare, we prefer to exclude it from the analysis and not to 
construct the extension of the random matrix theory. It could not
be tested convincingly on a single case, anyway.

Most importantly, we could not extend the random model fits to
higher $F_0$ values for a very simple
reason. There, we observed indications of the 
opening of other ionization channels
(see Eq.~(\ref{Gam})). A typical signature of such
a behaviour is the disappearance of the
singularity $\Gamma^{-1/2}$ in the distribution of the widths.
To understand this, note
that the typical Porter-Thomas distribution
behaves, for small widths, as 
$\Gamma^{M/2-1}$ with $M$ being
the number of open channels \cite{brody,dupret}. 
In the chaos assisted tunneling
process leading to ionization, while the full distribution
differs from the Porter-Thomas distribution, as exemplified
earlier for the single channel case, the small width functional behaviour is
similar in both cases.

The study of the available data reveals
that the opening of the second, and possibly third ionization
channel appears gradually with the increasing microwave amplitude,
$F_0$. Thus the different possible ionization channels are not
of equal importance, i.e., they are not equivalent (in the language
of the random matrix theory \cite{HILSS}). To build the random
matrix model of the process one then needs to introduce
additional free parameters describing the strength of the coupling
to the additional ionization channels, i.e. various values
for the $\gamma_k$ in Eq.~(\ref{Gam}). Although such a procedure
is quite straightforward, it is clear that fitting these parameters
to two data sets (shifts and widths) provides little information
and must be ambiguous. Typical distributions of the square root of
the width obtained for large microwave amplitudes are presented in
Fig.~\ref{fig9} for LPM and CPM wave packets.
Note the presence of the hole for small widths.

On the other hand, since in the perturbative limit, the level shifts
depend only on the real coupling between the localized state and
the remaining chaotic subspace, 
Eqs.~(\ref{s0ab}), (\ref{typical}), 
one can expect that the shifts will
be still well described by the Cauchy law, Eq.~(\ref{eqs}), independently
of the opening of additional ionization channels.
Indeed, it is the case as exemplified in
Fig.~\ref{fig9} for LPM and CPM wave packets.

Similarly, the opening of additional ionization channels is expected
in the semiclassical limit. The limit is realized by decreasing
the microwave frequency, then the wave packet is composed of
circular states of higher $n_0$. While the data corresponding to
the single channel decay have been obtained for $n_0=40$ at
$F_0=0.0426,$
we observed the opening of the second channel for the same $F_0$
starting at $n_0=60$. Panel (e) in Fig.~\ref{fig9}
presents the histogram of the square root of the width for $n_0=90$
and shows the existence of at least two open channels.
Again the corresponding shift distribution is not affected and
is well described by the Cauchy distribution [panel (b)].

\section{Physical Interpretation and Conclusions}

We have presented a statistical theory of ionization
catalyzed by chaos assisted tunneling. The corresponding
physical picture is built of a single state, localized on a stable
island and coupled (quantum mechanically, due to a finite value of
$\hbar$) to the surrounding chaotic sea. Once the tunneling
into the sea takes place, the diffusive 
chaotic excitation leads finally to
ionization.

A random matrix theory model allows us to determine analytically
the distribution of the energy shifts 
(induced by the interaction with the chaotic sea)
of the localized state, as well as
the distribution of its widths (ionization rates), in the
perturbative limit. Non-perturbative corrections may also be
understood and estimated. We concentrated on the simplest case of
single channel ionization -- the model then is characterized 
by few parameters only. In that case, both the distributions of shifts
and widths have long algebraic tails
explaining the large scale fluctuations of both quantities.
These fluctuations are a characteristic feature of chaos assisted
tunneling processes. Fluctuations (and universal properties of
fluctuations) are well established properties of chaotic systems.
In the ionization brought about by chaos assisted tunneling,
the combination of a weak tunneling process with chaotic
coupling to the continuum increases dramatically the range
of the fluctuations, by extending the distribution considerably 
towards extremely small widths, i.e. metastable states.

The developed theory has been confronted with numerical data
obtained for the shifts and widths of non-spreading wave packets --
states localized on a stable 1:1 resonance island between
the Kepler frequency of a Rydberg electron and the
frequency of an externally applied microwave field of either
linear or circular polarization
-- a system accessible to present experiments. The numerical
data have been obtained for simplified models of the atom -
a one-dimensional atom in LPM and a two-dimensional atom in
CPM. This allowed us to study the frequency range well in the
experimental region - the important atomic states building
the wave packet correspond to the principal quantum numbers
used in the experiments. The principal reason for the simplification
is that fully three-dimensional numerical calculations -- although
possible for a single set of parameters
as exemplified by us before \cite{zdb95} -- are
still prohibitive for present day computers.
More importantly, however, the statistical
properties of non-spreading wave packet states are not affected
by the 
reduced dimensionality of the atom as tested by us for the
three-dimensional CPM case.

It turns out that the developed statistical theory very well
describes the numerical data in the range of the single channel decay, 
which makes us confident that it contains
the essential ingredients of the physical process. The quality of the fits
allows to extract order from chaos, that is to extract from
strongly fluctuating quantities, see Fig.~\ref{fig1}, the physical
parameters describing the coupling of the non-spreading wave packet
to the chaotic states
(tunneling rate) and the 
ionization rate of chaotic states. 
These parameters exhibit reasonably smooth behaviour.
For example, we have shown
that secondary 
classical resonances inside the regular island increase the tunneling rate.
As an unambiguous signature of a
tunneling process
we also could demonstrate the exponential decrease of the tunneling rate
with the principal quantum number.

Let us emphasize the importance of the fluctuations of the ionization
rate (width) of non-spreading wave packets in the hydrogen atom.
In a real experiment, it is likely that the atoms will experience
various values of the microwave field strength -- either
because of spatial inhomogeneities or because they are prepared
by a slow increase of the microwave strength as explained in
\cite{zd:jpb97} -- and more or less
average the short range fluctuations of the ionization rate.
In the total, the residual ionization of the atom will be given
by the average ionization rate, a quantity dominated by the
fluctuations towards large ionization rates and which can be
 significantly larger than the typical ionization rate.
For example, for the data in Fig.~\ref{fig1} discussed
in this paper, the average ionization rate is about 6.4 times
larger than the typical ionization rate. In the limit
of the perturbative regime, the ratio of the two even diverges!
This is an example of physical processes as Levy flights where the
physics of a fluctuating system is dominated by rare events.

>From a practical point of view, the present study also tells us that
the lifetimes of the non-spreading wave packets either in CPM or in LPM
are rather long. Indeed, for $n_0=60$
and $F_0=0.0426$  -- these values are representative
of what could be used in a real experiment, microwave frequency
around 30 GHz and microwave field 
amplitude of the order of 10V/cm --
the typical lifetime of the non-spreading wave packet in CPM,
due to ionization catalyzed
by chaos assisted tunneling, is of the order of several micro-seconds, that is
about 100 000 Kepler periods. However, fluctuations
by one or two orders of magnitude are expected around this typical value.
Even the longest lifetimes should be shorter
than the natural lifetime, due to spontaneous emission, of the
order of a fraction of a second.  At higher $n_0\simeq 100,$
the typical ionization lifetime is of the order of several milliseconds,
i.e. 10 million Kepler periods, but still shorter than
the lifetime induced by spontaneous emission \cite{ibb97}. 
Hence, for practical
experiments in CPM, spontaneous emission should not be a problem. In LPM,
spontaneous emission is a slightly stronger effect, but 
largely dominated by chaos assisted tunneling ionization 
for $n_0\leq 100$
\cite{ab97}.

Finally, the physical situation and the model described
here are not restricted to atomic non-spreading wave packets.
It should describe physical systems where a given state is weakly
coupled to a dense family of completely different other states
which can decay on a rather long time scale. Then, the effective
decay rate of the initial state, induced by the coupling with
the family of decaying states, should present
huge fluctuations. An example is given in nuclear physics by the so-called
super-deformed nuclei \cite{superdeformed} where the ground state of a super-deformed
nucleus can only decay by coupling to highly excited
(hence chaotic) states of the non-deformed nucleus.
Our model then predicts the distribution of lifetimes
of super-deformed nuclei.

\acknowledgments
CPU time on a Cray C98 computer has been
provided by IDRIS and RZG.
Laboratoire Kastler Brossel de
l'Universit\'e Pierre
et Marie Curie et de l'Ecole Normale Sup\'erieure is
unit\'e associ\'ee 18 du CNRS. J.Z. acknowledges support of KBN
under project No.~2P03B~03810.
The additional support under the bilateral collaboration scheme
(J.Z. and D.D.) of the French Embassy in Poland, no.76209 
and the Programme International de Coop\'eration Scientifique
(CNRS) no.408 is appreciated.

\appendix
\section{Derivation of shift and width distributions}

We derive here the shift and width distributions for the random matrix model
described in section III. Starting from the Hamiltonian, Eq.~(\ref{mat1}),
we want to compute the complex eigenvalue close to zero. The real
part will be the desired shift and twice the imaginary part taken with 
minus sign the
desired width. 

We are interested in the case where the real coupling $\sigma$ and
the imaginary coupling $\gamma$ are sufficiently small for the
localized state not to be strongly mixed with the chaotic states. If this
condition is not fulfilled, the localized state
cannot be assigned
to a given eigenstate of the full ${\cal H}$ matrix and
the shift and width are ill-defined quantities.
In the following, we thus consider the perturbative limit where both
$\sigma$ and $\gamma$ are much smaller than $\Delta.$ Typically,
the localized state is then weakly coupled to the chaotic sea, itself
weakly coupled to the continuum. It may happen that the localized
state is accidentally almost degenerate with a chaotic eigenstate of
$H_0,$ bringing back the problem of assigning the strongly mixed state.
However, this is a rare event which - as it is shown below - affects
the tails of the distributions only. In the generic case, the couplings
are weak and the energy shift and the width can be calculated
perturbatively.

There are two small parameters for the perturbative analysis, namely
$\sigma/\Delta$ and $\gamma/\Delta.$ At first order in these small
parameters, there is no effect on the energy of the localized state
(no diagonal element). On the other hand, the localized state is contaminated
at first order in $\sigma/\Delta$ by the chaotic states.
The perturbed eigenstate $|\tilde{0}\rangle$ can be written
as:
\begin{equation}
|\tilde{0}\rangle = |0\rangle + \sigma \frac{1}{H_0} |V\rangle ,
\end{equation}
where $|0\rangle$ denotes the unperturbed localized state and
$|V\rangle $ is the $N$-component vector describing the coupling of the
localized state to the chaotic states. It can be also expanded on the
eigenstates $|\phi_i\rangle$ of $H_0$ with eigenvalues $E_i$ as:
\begin{equation}
|\tilde{0}\rangle = |0\rangle + \sigma \sum_{i=1,N}{\frac{\langle \phi_i
|V\rangle |\phi_i\rangle }{E_i}}.
\end{equation}

This admixture in the eigenstate results in a (real) energy shift at 
second order, i.e. proportional
to $\sigma^2$, given by:
\begin{equation}
s = - \sigma^2 \langle V \left| \frac{1}{H_0} \right| V\rangle =
- \sigma^2 \sum_{i=1}^N{\frac{|\langle \phi_i|V\rangle |^2}{E_i}}.
\label{rmts}
\end{equation}

Also, as the imaginary coupling $-i{\cal W}$, Eqs.~(\ref{mat1})
and (\ref{w}),
has non-zero diagonal elements, this implies a non-zero imaginary
part of the energy at order $\sigma^2\gamma,$ more precisely a width
given by:
\begin{equation}
w =  \gamma \sigma^2 \left| \langle W \left| \frac{1}{H_0} \right| V \rangle \right|^2
=    \gamma \sigma^2 \left| \sum_{i=1}^N{\frac{\langle W |\phi_i \rangle
\langle \phi_i|V\rangle }{E_i}} \right|^2.
\label{rmtw}
\end{equation}
Since $w$ appears as the square of a simpler quantity $x$, i.e.:
\begin{equation}
w = x^2
\end{equation}
with $x$ given by:
\begin{equation}
x= \sqrt{\gamma} \sigma \langle W \left| \frac{1}{H_0} \right| V \rangle
= \sqrt{\gamma} \sigma \sum_{i=1,N}{\frac{\langle W |\phi_i \rangle
\langle \phi_i|V\rangle }{E_i}},
\label{rmtx}
\end{equation}
we will consider in the following the statistical distribution of $x$
rather than $w.$

To obtain the statistical distributions within
our random matrix model, one has to average over the random ensemble,
i.e. over the various Gaussian random variables: the $N$ components
of $|V\rangle$, the $N$ components of $|W\rangle$ and the $N(N+1)/2$ 
independent matrix elements of $H_0.$ Because of the orthogonal invariance
of $H_0,$ the averaging over $|V\rangle$ and $|W\rangle$ is straightforward.
Thus, the distribution of shift is essentially the distribution of diagonal
elements of $1/H_0$ while the distribution of $x$ (square root of the width)
is essentially the distribution of matrix element of $1/H_0$ between
two statistically independent vectors. 

Since we are interested in the situation 
where a large number of chaotic states are coupled to the localized state,
we will take the limit $N \rightarrow \infty,$ keeping the mean level
spacing equal to a constant $\Delta$ and keeping $\gamma$ and $\sigma$ fixed,
such that the average coupling between a chaotic state and the localized state
is independent of $N.$ Then, in the sums in Eqs.~(\ref{rmts}) and (\ref{rmtx}),
there are more and more terms which contribute as $N$ is increased, but
corresponding to larger and larger energy denominators $1/E_i$ so that
the total sum has a well defined limit as $N \rightarrow \infty.$
Yet, in this limit, the scalar product $\langle W | V \rangle = \sum
{\langle W | \phi_i\rangle \langle \phi_i|V\rangle }$ appears as the sum
of the product of independent Gaussian variables which typically averages to
a small quantity. In other words, in the $N \rightarrow \infty$ limit,
$|V\rangle$ and $|W\rangle$ appear as independent orthogonal vectors,
so that the distribution of $x$ values is essentially the distribution
of non-diagonal elements of $1/H_0.$ Corrections due to non exact orthogonality
will modify the distribution at order $1/N$ only.

The calculation of the sums in Eqs.~(\ref{rmts}) and (\ref{rmtx}) is not
completely straightforward, because the various energies in the denominators
are {\em correlated} if $H_0$ is a GOE random matrix. Before studying
this case, let us consider the simpler case where the spectrum of $H_0$
is a set of {\em uncorrelated} eigenvalues, i.e. a Poisson spectrum.
Then, all quantities in the numerators are uncorrelated 
random Gaussian variables and denominators are uncorrelated energies
with mean density $1/\Delta.$ In this case, the calculation
can be done exactly as shown in Refs.~\cite{LU96,mello} and the distribution
is a Cauchy distribution whose half-width is proportional
to the average absolute value of the numerator, that is:
\begin{equation}
P_{\rm Poisson}(s)= \frac{1}{\pi} \frac{s_0}{s_0^2+s^2}
\end{equation}
with
\begin{equation}
s_0=\pi \frac{\sigma^2 \ \overline{|\langle \phi_i|V\rangle |^2}}{\Delta}
\end{equation}
where the bar denotes the average value over the random matrix ensemble.
Here, it is simply the variance of the components of $|V\rangle $,
that is 1 in our model (see section III). Hence,
\begin{equation}
s_0= \frac{\pi \sigma^2}{\Delta}.
\label{s0}
\end{equation}

Similarly, one gets
\begin{equation}
P_{\rm Poisson}(x)= \frac{1}{\pi} \frac{x_0}{x_0^2+x^2}
\label{pxp}
\end{equation}
with
\begin{equation}
x_0 = \pi \frac{\sigma \sqrt{\gamma} 
\ \overline{|\langle W |\phi_i \rangle \langle \phi_i|V\rangle |}}{\Delta}.
\end{equation}
The average over the random ensemble (Gaussian integral) 
gives a $2/\pi$ factor, resulting in:
\begin{equation}
x_0= \frac{2\sigma \sqrt{\gamma}}{\Delta}.
\end{equation}
For the width itself, we obtain the distribution given in
section III, Eq.~(\ref{eqx}). 

We now turn to the random matrix case, where $H_0$ 
is a standard real symmetric random
matrix belonging to the GOE (Gaussian Orthogonal Ensemble).
In Ref.~\cite{brouwer}, P.~Brouwer introduced a slightly different class
of random matrices, namely the Lorentzian Orthogonal Ensemble ($\Lambda$OE)
which has the following Lorentzian probability distribution:
\begin{equation}
P(H)= \left(\frac{\lambda^2}{\pi}\right)^{N(N+1)/4}
\prod_{i=1,N}{\frac{\Gamma(i)}{\Gamma(i/2)}}
\ \ \det \left[ (\lambda^2+H^2)^{-(N+1)/2}\right ]
\label{loe}
\end{equation}
where $\lambda$ is a parameter describing the width of the distribution.

As shown in Ref.~\cite{brouwer}, although the $\Lambda$OE has different global
statistical properties (as e.g. density of states) than the GOE, it has
locally the same joint probability distribution function of the eigenvalues
and consequently the same spacing distribution, the same short range
correlation functions, etc. The mean level spacing close to the
centre of the spectrum (energy equal to zero) is:
\begin{equation}
\Delta= \frac{\lambda \pi}{N}
\end{equation}
Hence, to calculate the shift and the width distribution, we can replace
$H_0$ by a random matrix of the $\Lambda$OE. 
The $\Lambda$OE has the nice property \cite{brouwer}
that if $H_0$ is distributed according to $\Lambda$OE, then $1/H_0$ is also
distributed according to $\Lambda$OE with width $1/\lambda.$
Morevoer, if $H_0$ is distributed according to $\Lambda$OE, then 
every submatrix of $H_0$ is also distributed according to $\Lambda$OE,
with the same width \cite{brouwer}.
The price to pay for such nice properties is that one looses the 
statistical independence of the various matrix elements valid for the GOE, 
i.e. $P(H)$ cannot
be factorized as a product of distributions of elementary matrix elements.
However, this is not a problem for the quantities we are interested in.
The distribution of shifts is obtained straightforwardly, as it is a
diagonal element of $1/H_0$, hence a $1\times 1$ submatrix of $1/H_0$,
which - from the two properties just described - is given by
Eq.~(\ref{loe}) for $N=1$ and $\lambda=\pi/N\Delta.$
The result is exactly equal to the Poisson result, i.e. the Cauchy
(or Lorentzian) distribution of Eq.~(\ref{s0}).

For the width, the situation is slightly more complicated as we need
to know the distribution of a non-diagonal element of $1/H_0.$ 
The same trick works, but we have now to extract for the $\Lambda$OE matrix
$1/H_0$ a $2\times 2$ submatrix and consider the distribution
of non-diagonal element, that is to average over the 2 diagonal elements:
\begin{equation}
P(H_{12}) = \int\!\!\int{P(H)\ dH_{11}\ dH_{22}}
\end{equation}
where $P(H)$ is given by Eq.~(\ref{loe}) for $N=2$ and $\lambda=\pi/N\Delta.$
The integral over diagonal elements is trivial. The result is the following
distribution for the square root of the width:
\begin{equation}
P_{\rm GOE}(x) = \frac{2x_0'}{\pi^2 (x_0'^2+x^2)} \left[
1+\frac{{\rm arcsinh}(x/x_0')\ x_0'^2}{x\sqrt{x_0'^2+x^2}}\right ]
\label{px}
\end{equation}
with
\begin{equation}
x_0' = \frac{\pi \sigma \sqrt{\gamma}}{\Delta}
\end{equation}
For the width itself, we obtain the distribution given in
section III, Eq.~(\ref{eqx2}).

The GOE distribution, Eq.~(\ref{px}) has the same behaviour
as the Cauchy distribution obtained for the Poisson ensemble, Eq.~(\ref{pxp}),
that is a constant value near $x=0$ followed by a $1/x^2$ decrease at
large distance. In fact, the two distributions are very similar
with slightly different widths and are almost impossible to distinguish
by eye.

The distributions obtained in the various cases agree exactly with
numerically calculated distributions from a large number of realizations
of the Poisson and GOE random ensembles.

Note that the distributions obtained here
have long tails with an algebraic decay as $s$ (or $x$) tends to infinity.
This is unusual for quantities which are sums of statistically uncorrelated
individual terms and - as a consequence of the central limit theorem - 
have usually Gaussian distributions. The reason is that the central limit
theorem cannot be used here because the variance of each individual term
in infinite, see Ref.~\cite{LU96,mello}. 
The latter property is due to the $1/E_i$ dependence which
decays only slowly for large $E_i.$ 

On a double logarithmic scale, see Fig.~\ref{fig2}, the distribution of shifts
shows two different regimes: constant near the origin, and an asymptotic
$1/s^2$ behaviour for large shifts. The cross-over between the two regimes
is for $s=s_0;$ the latter value, Eq.~(\ref{s0}), corresponds to the typical
shift due to a chaotic level lying at a distance $\Delta$ from the localized
state, i.e. to the typical shift due to the nearest state. Hence, large
energy shifts $s\gg s_0$ are due to situations where one chaotic level
is much closer in energy than $\Delta.$ 
In such a case, one term is dominant in the sum, Eq.~(\ref{rmts}), a two level
approximation can be used, which produces the correct $1/s^2$ behaviour.
On the other hand, the ``constant" regime, $s\ll s_0$ corresponds to
situations where the various terms in Eq.~(\ref{rmts}) interfere destructively,
giving a total sum typically smaller than the largest individual terms: this
is an intrisically ``multi-level" situation where quantum destructive 
interferences play an important role.
Exactly the same thing takes place for the distribution of widths.
The largest widths - in the $w^{-3/2}$ regime - are obtained when a single
level dominates the sum, Eq.~(\ref{rmtx}), while the $w^{-1/2}$ regime
correspond to the ``multi-level" situation.

Finally, let us discuss what happens when the perturbative approach
breaks down. This takes place when one chaotic level is very close to
the localized level, closer than their average coupling $\sigma.$ 
There, the strong mixing between states invalidate the expressions
for the shift, Eq.~(\ref{rmts}) and the width, Eq.~(\ref{rmtx}).
The actual shift and width do not diverge, in contrast to the perturbative
expressions. This means that the actual distribution cannot have
an algebraic tail towards infinity, but should show a cut-off when
perturbation theory breaks down. As explained above, this takes place
for one of the $E_i$ of the order of $\sigma$, corresponding to:
\begin{eqnarray}
s_{\rm cutoff}\simeq \sigma\\
w_{\rm cutoff}\simeq \gamma
\end{eqnarray}
in agreement with our numerical observations.

\begin{figure}
\caption{Typical fluctuations of the width (ionization rate)
and the energy shift (with respect to its unperturbed position)
of the non-spreading wave packet of a hydrogen atom exposed to a
circularly polarized microwave field. The wave-packet
is a stationary state of the atom dressed by the microwave field and at the
same time a wave packet orbiting around the nucleus at exactly
the microwave frequency, without spreading. In the language of non-linear
physics, the wave packet is at the center of the primary resonance
island between the microwave frequency and the internal Kepler frequency of the
electron. It is coupled to the surrounding chaotic states by tunneling
and therefore has an ionization rate induced by a
chaos assisted tunneling process which shows huge fluctuations
when a parameter is varied. The data presented
are obtained for small variations
of the effective principal quantum number $n_0$ around 40,
a scaled
microwave electric field $F_0=Fn_0^4=0.0426$ and a microwave
frequency $\omega = 1/n_0^3$. To present both
plots on the logarithmic scale 
(more convenient to show the fluctuations over
several orders of magnitude) we plot the absolute value of the shift
rather than
the shift itself.
 }
\label{fig1}
\end{figure}

\begin{figure}
\caption{Distributions of the energy shifts (a) and widths (b) of the
non-spreading wave packet of a hydrogen atom in a circularly polarized microwave
field. The data are those of
Fig.~\protect{\ref{fig1}}. They are plotted on a double
logarithmic scale which clearly shows the large fluctuations.
Both the shift (actually the absolute value of the shift) and the
width have been rescaled to their typical values, as explained
in the text.}
\label{fig2}
\end{figure}

\begin{figure}
\caption{Comparison between the distributions of the energy shifts
of the non-spreading wave packet of a hydrogen atom
in circularly polarized microwave
(large bins histogram)
with the distribution
obtained for our Random matrix model (small bins histogram).
The data are those of
Fig.~\protect{\ref{fig1}} with the same rescaling.
(a) Linear scale; the agreement is excellent. The two
histograms follow exactly a Cauchy distribution, Eq.~(\protect{\ref{eqs}}).
(b) Double
logarithmic scale which emphasizes the long tail of the
distributions. Again, the agreement is remarkable. The dashed
line is the pure Cauchy distribution predicted in the perturbative
regime and differs from the numerical result at
large shifts.}
\label{fig2.1}
\end{figure}

\begin{figure}
\caption{Same as Fig.~\protect{\ref{fig2.1}}, but for the width.
(a) Linear scale for the distribution of the square root
of the width. The dashed line is the analytical result (Cauchy distribution),
Eq.~(\protect{\ref{eqx}}),
for a Poisson spectrum and the continuous solid line the
analytical result, Eq.~(\protect{\ref{eqx2}}), for a GOE spectrum. 
The agreement
with the Poisson prediction is slightly better, although all distributions
are very similar. (b) Double logarithmic scale showing the agreement
of the Random Matrix model with the real system, even for the tail
of the distribution. The dashed line is the analytical perturbative
expression, Eq.~(\protect{\ref{eqx}}), which differs from the numerical result
in the tail. The dotted line is the Porter-Thomas distribution,
Eq.~(\protect{\ref{PT}}), for a system without chaos assisted
tunneling; it describes the exponentially small tail of the
distribution.}
\label{fig2.2}
\end{figure}

\begin{figure}
\caption{
Typical shift (with respect to the 
unperturbed energy level)
and width (ionization rate) of the non-spreading wave packet
of the hydrogen atom in circularly polarized microwave.
Each point in the plot is extracted from the analysis of a distribution
similar to the one in Fig.~\protect{\ref{fig2}} built
from several hundred independent diagonalizations of the Hamiltonian,
Eq.~(\protect{\ref{hcirc}}), at neighbouring values of the
field strength and frequency, around $n_0=40.$
The statistical analysis of
each distribution is done as in Fig.~\protect{\ref{fig2.1}}
and Fig.~\protect{\ref{fig2.2}}.
The typical values of the energy shift (circles) and width (squares)
globally increase with the microwave field strength, but with
bumps which are obviously correlated.
The long-dashed line represents the mean energy 
level spacing $\Delta$ between
chaotic levels
(all quantities are plotted in atomic units). The typical width is smaller
than the typical shift, itself smaller than $\Delta,$ which proves that
the data are obtained in the perturbative regime where
chaos assisted tunneling is only a small perturbation. For details, see text.
}
\label{fig3}
\end{figure}

\begin{figure}
\caption{The non-spreading
wave packet of a hydrogen atom in circularly polarized microwave. Tunneling rate 
to the surrounding chaotic sea $\sigma/\Delta$ (a)
and chaotic ionization rate $\gamma/\Delta$ (b) of the chaotic states
as extracted from the data in Fig.~\protect{\ref{fig3}}, using our
simple statistical model based on Random Matrix Theory.
These two quantities are dimensionless (rescaled to the
mean level spacing) and smaller than 1 in the perturbative regime.
The tunneling rate has large oscillations when the scaled
microwave field is varied, as a consequence of secondary resonances
occuring in the primary island where the wave packet is localized.
The chaotic ionization rate increases very rapidly at low $F_0$,
as a consequence of the destruction of barriers slowing down
the chaotic diffusion towards ionization, and further saturates
at a rather constant value when chaotic dynamics is reached.
}
\label{fig4}
\end{figure}

\begin{figure}
\caption{Plots of the typical shift and width of the non-spreading wave packet,
as in Fig.~\protect{\ref{fig3}}, for
a hydrogen atom in linearly polarized microwave field, around $n_0=40.$
Again, the
typical values of the energy shift (circles) and width (squares)
globally increase with the microwave field strength, 
with
bumps which are obviously correlated.
The long-dashed line represents the mean energy spacing $\Delta$ between
chaotic levels
(all quantities are plotted in atomic units). The typical width is smaller
than the typical shift, itself smaller than $\Delta,$ which proves that
the data are obtained in the perturbative regime where
chaos assisted tunneling is only a small perturbation.
}
\label{fig5}
\end{figure}

\begin{figure}
\caption{Same as Fig.~\protect{\ref{fig4}}, but for a hydrogen atom in linearly polarized microwave.
Again, the bumps in $\sigma/\Delta$ (a) are related to the secondary
resonances in the system while $\gamma/\Delta$ (b) is more or less constant
as soon as the chaotic regime is reached.
}
\label{fig6}
\end{figure}

\begin{figure}
\caption{Same as Fig.~\protect{\ref{fig3}}, but plotted for
fixed classical dynamics (fixed $F_0=0.0426$) as a function of the effective
principal quantum number $n_0=1/\hbar_{\rm eff}.$ Because
the primary resonance island has a fixed structure, the bumps
visible in Fig.~\protect{\ref{fig3}} are almost absent. Both
the typical shift and typical width decrease exponentially with $n_0$,
a signature of a tunneling process. The long-dashed curve shows the
mean energy spacing $\Delta$ between chaotic states.
}
\label{fig7}
\end{figure}

\begin{figure}
\caption{Same as Fig.~\protect{\ref{fig4}}, but plotted for
fixed classical dynamics (fixed $F_0=0.0426$) as a function of the effective
principal quantum number $n_0=1/\hbar_{\rm eff}.$
The tunneling rate $\sigma/\Delta$ (a) decreases exponentially with $n_0$ 
(note
the logarithmic scale)
which proves that the process involved is actually tunneling. From the rate
of the exponential decrease we are able to extract the ``tunneling action"
$S=0.06 \pm 0.01.$
The chaotic ionization rate $\gamma/\Delta$ (b) smoothly and slowly
evolves with $n_0$ (note the linear scale), approximately as $n_0^2.$
}
\label{fig8}
\end{figure}

\begin{figure}
\caption{Dimensionless shift distribution  (a-c) and distribution of the square
root of the dimensionless width (d-f)
for the non-spreading wave packet
of a hydrogen atom in a microwave field.
The distributions of the square root of the width
deviate from our simple
statistical model which predicts a Cauchy law, because
several ionization channels are opened at high field strength or for
large $n_0$.
On the contrary, as predicted by the model, the distributions
of shifts remain close to Cauchy distributions (solid lines in panels a,b,c)
and are not sensitive to the number of
open channels.
Panel (a) and (d) correspond to circularly polarized microwave field,
$n_0=40$, $F_0=0.068$, panels (b) and (e) to circularly polarized microwave
field,
$F_0=0.045$, but $n_0=90$, i.e., 
for a much smaller 
average frequency $\omega=1/n_0^3$,
that is deeper in the semiclassical regime. Panels (c) and (f)
correspond to linearly polarized microwave field,
$n_0=40$, $F_0=0.076.$
}
\label{fig9}
\end{figure}

\begin{references}
\bibitem{comment} $n_0$ is defined from the microwave frequency
through Eq.~(\protect\ref{n0}) and is not necessarily an integer.
Therefore, we identify the wave packet by its large overlap with
the hydrogenic state $|n\rangle$ with n the closest integer to $n_0.$
\end{references}
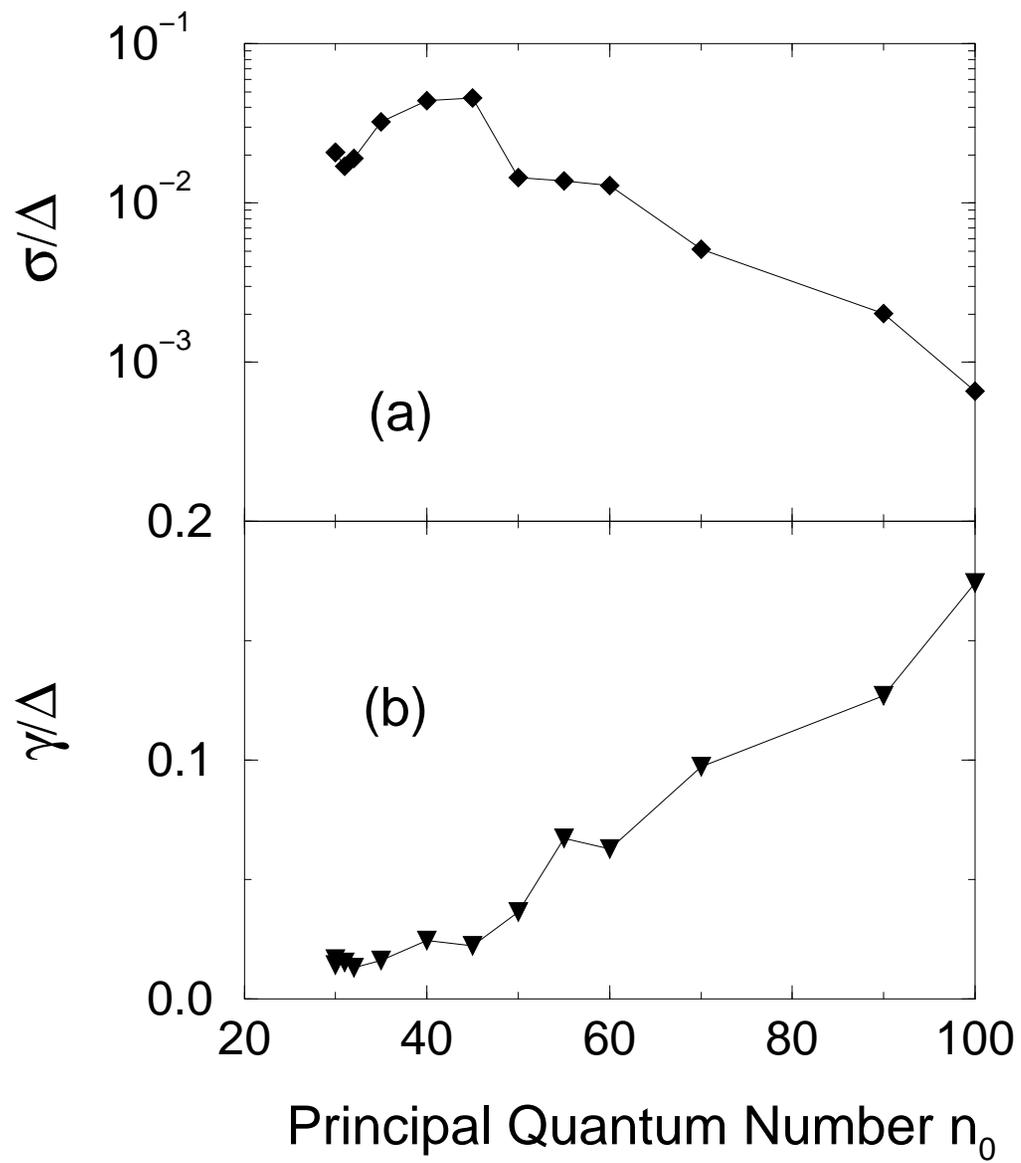
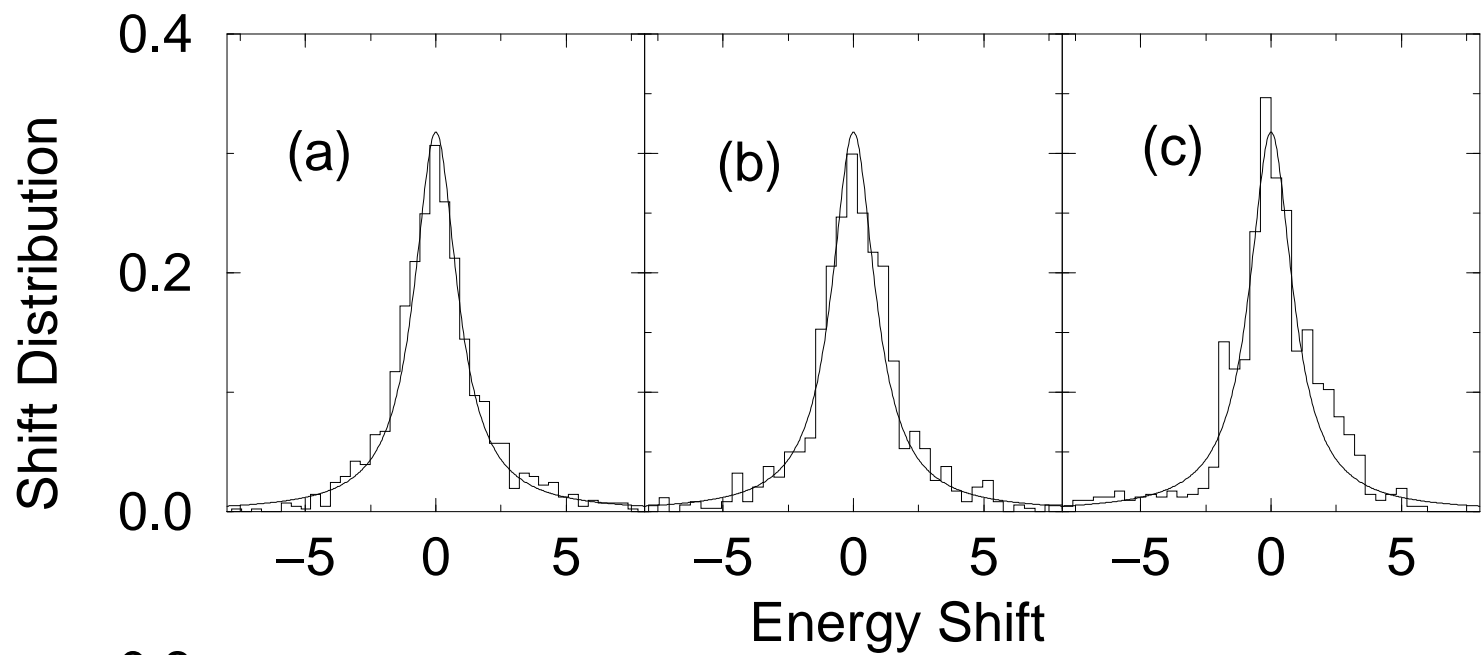
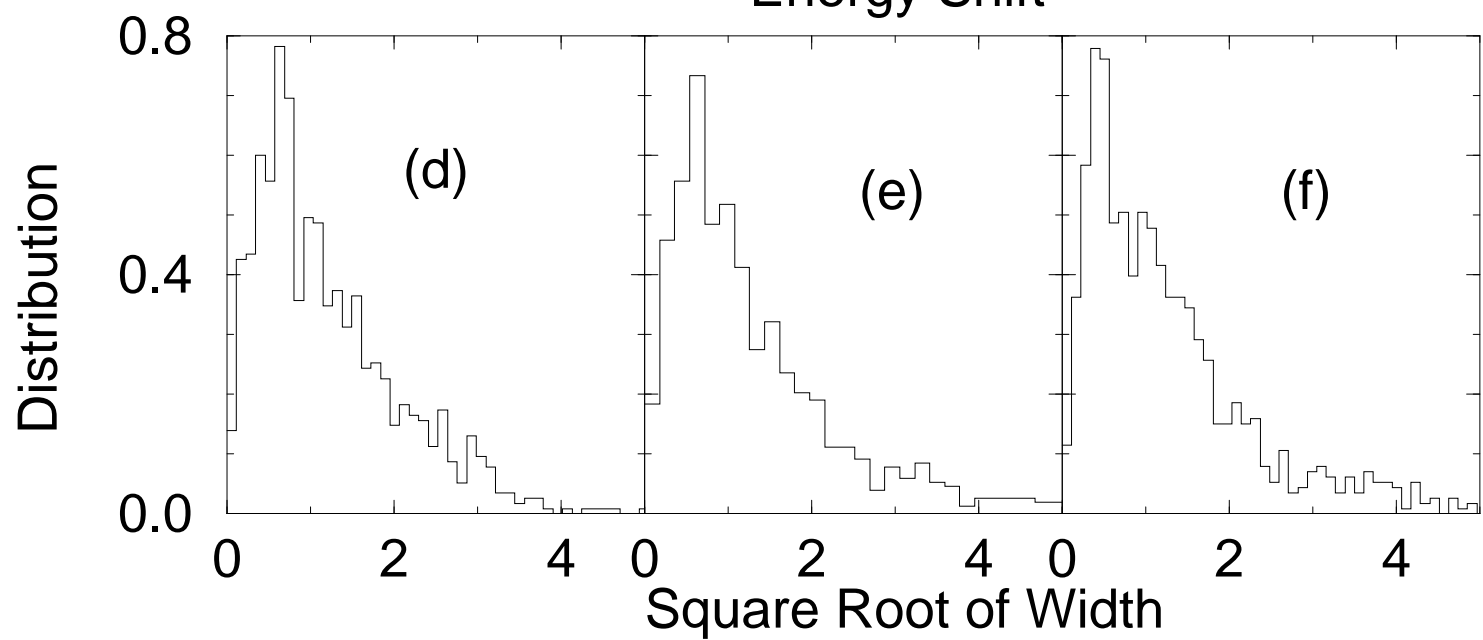
\end{document}